\documentclass[aps,preprint]{revtex4}%
\usepackage{amsfonts}
\usepackage{amsmath}
\usepackage{amssymb}
\usepackage{graphicx}
\usepackage{pdfpages}
\usepackage{float}
\usepackage{mathrsfs}
\usepackage{color}
\usepackage{xcolor}
\usepackage{caption}
\usepackage{appendix}
\usepackage{lineno,hyperref}
\usepackage{changes}%
\hypersetup{colorlinks =true, allcolors = blue}
\providecommand{\U}[1]{\protect\rule{.1in}{.1in}}

\begin{document}
\begin{center}
{\Large Quazinormal modes and greybody factor of black hole
surrounded by a quintessence in the S-V-T modified gravity as well as shadow}

Ahmad Al-Badawi 

Department of Physics, Al-Hussein Bin Talal University, P. O. Box: 20,
71111, Ma'an, Jordan

\bigskip E-mail: ahmadbadawi@ahu.edu.jo, \\
{\Large Abstract}
\end{center}
The purpose of this study is to investigate the quasinormal modes (QNMs), greybody factors (GFs) and shadows in a plasma of a black hole (BH) surrounded by an exotic fluid of quintessence type in a scalar-vector-tensor modified gravity. The effects of a quintessence scalar field and the modified gravity (MOG) field on the QNM, GF, and shadow are examined. Using the  sixth-order WKB approach, we investigate the QNMs of massless scalar and electromagnetic perturbations.  Our findings show that as the quintessence and the MOG parameter ($\epsilon$ and $\alpha$) increase, the oscillation frequencies decrease significantly. Gravitational wave damping, on the other hand, decreases with increasing $\epsilon$ and $\alpha$. In addition, we obtain an analytical solution for the transmission coefficients (GF) and demonstrate that more thermal radiation reaches the observer at spatial infinity as both the $\epsilon$ and $\alpha$ parameters increase. We also investigate the effect of the plasma background on the BH shadow and show that as the plasma background parameter increases, the shadow radius slightly shrinks. Nevertheless, the shadow radius increases as  $\alpha$ and $\epsilon$ increase. Particularly intriguing is the fact that increasing $\epsilon$ has a greater impact on the shadow radius than increasing  $\alpha$, indicating that the quintessence parameter has a greater impact than the MOG parameter.

\section{Introduction}

Observations of magnitude-Redshift relationships of astronomical objects suggest that our universe is expanding rapidly \cite{ex1,ex3}. Even though Einstein's cosmological constant was designed to explain this phenomenon, its theoretical value is quite different from the experimental results. This phenomenon has thus prompted physicists to seek new explanations. A popular proposal states that cosmological observations suggest the existence of a special type of energy that pervades the entire universe, known as dark energy, which could be the cause of negative pressure \cite{68}. According to the standard cosmological model, our current universe is made up $73\%$ dark energy of the total energy density. The quintessence, whose state equation is given by, has attracted a lot of attention as one of the most suitable dark energy candidates. There are two proposed theories for dark energy: quintessence and the cosmological constant. The quintessence is characterised by the equation of state   ($\gamma =$ Pressure/energy density). The study of Kiselev \cite{kiselev} examined Einstein's field equations to describe a BH surrounded by quintessential matter. He derived a solution based on the state parameter $\gamma$ that characterizes the quintessence through his analysis. Since then, various investigations have been undertaken on the features of the BH with quintessence, with substantial results \cite{q1,q2,q3,q4,q5,q6,q7,q8,q9,q10,q11,q12,q13,q14,q15}.\\ On the other hand,  General Relativity (GR) defines BHs as regions in spacetime where classical physics fails at the BH's intrinsic singularity. Although GR has been successful, it has faults. The theory has two major flaws: singularities \cite{qn5,qn6} and the lack of observational evidence for dark matter \cite{qn7}.   Considering this, it is either necessary to modify the theory of GR or to claim dark matter exists \cite{qn8,qn9,qn10,qn11,qn12}. To investigate the nature of BHs, we must first develop a theory that eliminates the ambiguities mentioned above. One of the successful modified gravity theory is the scalar-vector-tensor (S-V-T), also known as MOG \cite{bq37}. This theoretical model accurately describes Solar System observations \cite{bq38}, galaxies' rotation curves \cite{bq39}, galaxy cluster dynamics \cite{bq40}, and predictions for the Cosmic Microwave Background \cite{bq41}.  Recent research suggests that BH solutions in the MOG gravity model can accurately reproduce GR's exact solutions \cite{bq42,bq43}.  Following the development of the physical solution for a MOG BH,  the authors of Ref. \cite{bq44} studied the astrophysical consequences of the existence of a quintessence scalar field. They consider a MOG BH with quintessence (QMOG BH) \cite{bq44} and investigate the impact of quintessence on the formation of horizons, equations of motion, effective potential, and astrophysical implications such as deflection angle, shadow, and Shapiro time delay. \\ In this paper, we look at the quasinormal modes (QNMs), greybody factors (GFs), and shadows in the presence of plasma, as well as the effects of a quintessence scalar field in the QMOG spacetime. The study is divided into five sections (Sections II-V), in which we first describe the QMOG BH spacetime, then analyse the effective potentials, and last compute and analyse the complex quazinormal frequencies for scalar and electromagnetic fields. In addition, we analyse the GFs associated with QMOG BH,  the shadow in the presence of a plasma background as well as its effect on the shadow itself.  Section 6 concludes with a summary and discussion.

\section{ Review of QMOG spacetime}

The action in the STVG theory is given by \cite{bq37}:
\begin{equation}
S=\frac{1}{16\pi G_{N}\left( 1+\alpha \right) }\int d^{4}x\sqrt{-g}\left( R-%
\frac{1}{4}B^{\mu \nu }B_{\mu \nu }\right)\, \label{ac1}
\end{equation}%
where $\alpha$ is a dimensionless parameter
and $ G_{N}$ is the Newtonian
constant. When varying the action (\ref{ac1}) with respect to metric tensor, we can obtain the following field equations for the matter-free MOG BH metric spacetime \cite{bq37}:
\begin{equation*}
R_{\mu \nu }=8\pi G_{N}\left( 1+\alpha \right)
T_{\mu\nu}^{(\phi)},
\end{equation*} \begin{equation*}
\frac{1}{\sqrt{-g}}\partial _{\nu }\left( \sqrt{-g}B^{\mu \nu }\right) =0,
     \end{equation*}
     \begin{equation} 
\partial _{\sigma }B_{\mu \nu }+\partial _{\mu }B_{\nu \sigma }+\partial
_{\nu }B_{\sigma \mu }=0.
     \end{equation}   The energy-momentum tensor is
\begin{equation}
T_{\mu\nu}^{((\phi))}=-\frac{1}{4\pi }\left( B_{\mu \alpha }B^{\nu \alpha }-\frac{1%
}{4}\delta _{\mu }^{\nu }B^{\alpha \beta }B_{\alpha \beta }\right) .
\end{equation}
 Through the decomposition of the Einstein field equation's energy momentum tensor into two components  \cite{kiselev}
\begin{equation}
     T_{\mu\nu}=T_{\mu\nu}^{q}+T_{\mu\nu}^{(\phi)}   \end{equation}   
     where $T_{\mu\nu}^{q}$  for the quintessence part.
Finally, the static MOG spacetime  solution in the presence of a quintessence can be written as \cite{bq42,bq43,kiselev,bq44}
\begin{equation}
ds^{2}=-f\left( r\right) dt^{2}+\frac{1}{f(r)}dr^{2}+r^{2}\left( d\theta
^{2}+\sin ^{2}\theta d\phi ^{2}\right),  \label{M1}
\end{equation}%
where  
\begin{equation}
f\left( r\right) =1-\frac{2G_{N}\left( 1+\alpha \right) M}{r}+\frac{\left(
1+\alpha \right) G_{N}Q^{2}}{r^{2}}-\frac{\epsilon\left( 1+\alpha \right) G_{N}}{%
r^{3\gamma +1}},  \label{mf1}
\end{equation}%
where $M$ is the BH mass, $Q=\sqrt{\alpha G_{N}}M$ is the effective
(non-electric) charge, $G_{N}$ Newton's constant, while $\alpha $ is the
enhancement (MOG) parameter, $\epsilon$ is a normalization factor and $\gamma$ is the
state parameter of the quintessence ($-1<\gamma <-1/3$).
In the following we will call the BH described by the metric (\ref{M1}) as
the QMOG BH \cite{bq44}. BH metrics for QMOG reduce to the following geometries in the following cases: 
\begin{center}
    \begin{tabular}{|c|c|c|} \hline 
         & $f\left( r\right) $ & Geometry 
\\ \hline
$\gamma =-1$ & $1-\frac{2G_{N}\left( 1+\alpha \right) M}{r}+\frac{\left(
1+\alpha \right) G_{N}Q^{2}}{r^{2}}-\epsilon \left( 1+\alpha \right)
G_{N}r^{2}$ & anti de-Sitter Schwarzschild \\ 
$\gamma =-2/3$ & $1-\frac{2G_{N}\left( 1+\alpha \right) M}{r}+\frac{\left(
1+\alpha \right) G_{N}Q^{2}}{r^{2}}-\epsilon \left( 1+\alpha \right) G_{N}r$
& effective RN in SVTQ  \\ 
$\epsilon =0$ & $1-\frac{2G_{N}\left( 1+\alpha \right) M}{r}+\frac{\left(
1+\alpha \right) G_{N}Q^{2}}{r^{2}}$ & effective RN in SVT  \\ 
$\epsilon =0,Q=0$ & $1-\frac{2G_{N}\left( 1+\alpha \right) M}{r}$ & 
Schwarzschild in SVT%
\\ 
\hline 
\end{tabular}
\end{center} where RN = Reissner-Nordstr\"{o}m. \\
To appropriately examine the QMOG BH's singularity and uniqueness, we must offer a scalar-invariant analysis:
\begin{equation}
R=\frac{3\epsilon \gamma \left( 3\gamma -1\right) \left( 1+\alpha \right) }{%
r^{3\left( \gamma +1\right) }},
\end{equation}%
\begin{equation}
R_{\mu \nu }R^{\mu \nu }=\frac{\left( 1+\alpha \right) ^{2}\left( 9\epsilon
^{2}r^{2}\gamma ^{2}\left( 5+6\gamma +9\gamma ^{2}\right) -36\epsilon
M^{2}r^{1+3\gamma }\gamma \left( 1+\gamma \right) \alpha +8M^{4}r^{6\gamma
}\alpha ^{2}\right) }{r^{8+6\gamma }},
\end{equation}%
\begin{equation}
R_{\mu \nu \alpha \beta }R^{\mu \nu \alpha \beta }=\frac{\left( 1+\alpha
\right) ^{2}\left( 3\epsilon ^{2}r^{2}\left( 4+\gamma \left( 20+3\gamma
\left( 17+9\gamma \left( 2+\gamma \right) \right) \right) \right) \right) }{%
r^{8+6\gamma }}+
\end{equation}%
\[
\frac{\left( 1+\alpha \right) ^{2}\left( \left( 12\epsilon Mr^{1+3\gamma
}\left( 1+\gamma \right) \left( r\left( 4+6\gamma \right) -M\left( 4+9\gamma
\right) \alpha \right) \right) +8M^{2}r^{6\gamma }\left( 6r^{2}-12Mr\alpha
+7M^{2}\alpha ^{2}\right) \right) }{2r^{8+6\gamma }}.
\]
It should be noted that the QMOG spacetime has a physical singularity at $r = 0$.

\begin{figure}
    \centering
\includegraphics{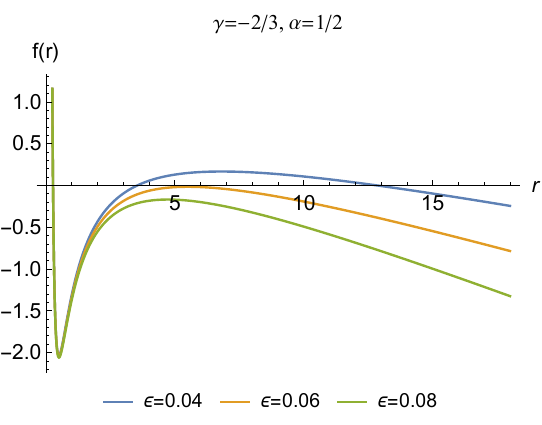}
    \caption{The plot of metric function (\ref{mf1}) for $M=1$ and $G_{N}=1$.}
    \label{fig1}
\end{figure}
In the absence of quintessence, the spacetime metric (\ref{M1}) can have at most two horizons depending on the parameter, $\alpha$ \cite{bq42}. In the presence of quintessence, a cosmological horizon appears. Accordingly, the QMOG BH metric   can have three horizons as shown in Fig. \ref{fig1} (see \cite{bq44} for details). The Figure  illustrates how the number of horizons in the QMOG BH decreases as $\epsilon$ increases from three to two or even one.

\section{QNMs}
\begin{figure}
    \centering
{{\includegraphics[width=7.5cm]{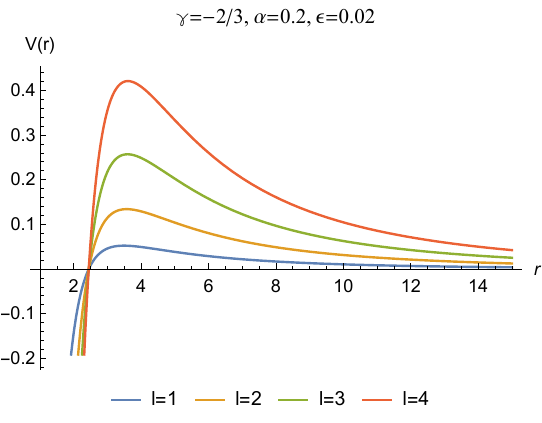} }}\qquad
    {{\includegraphics[width=7.5cm]{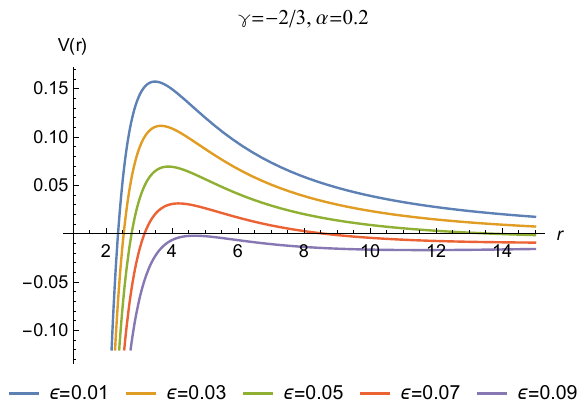}}}
    \caption{Scalar potentials (\ref{pot1})
 with different values of $l$ (left) and $\epsilon$ (right) ($M=1$).}
    \label{fig2}
\end{figure}

\begin{figure}
    \centering
    \includegraphics{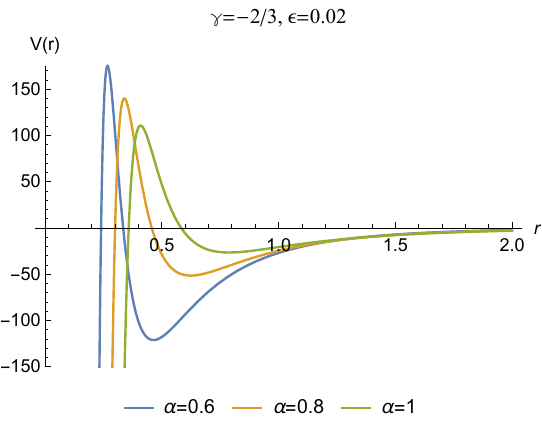}
    \caption{Scalar potentials (\ref{pot1})
 with different values of $\alpha$ ($M=1$).}
    \label{fig3}
\end{figure}

\begin{figure}
    \centering
{{\includegraphics[width=16cm]{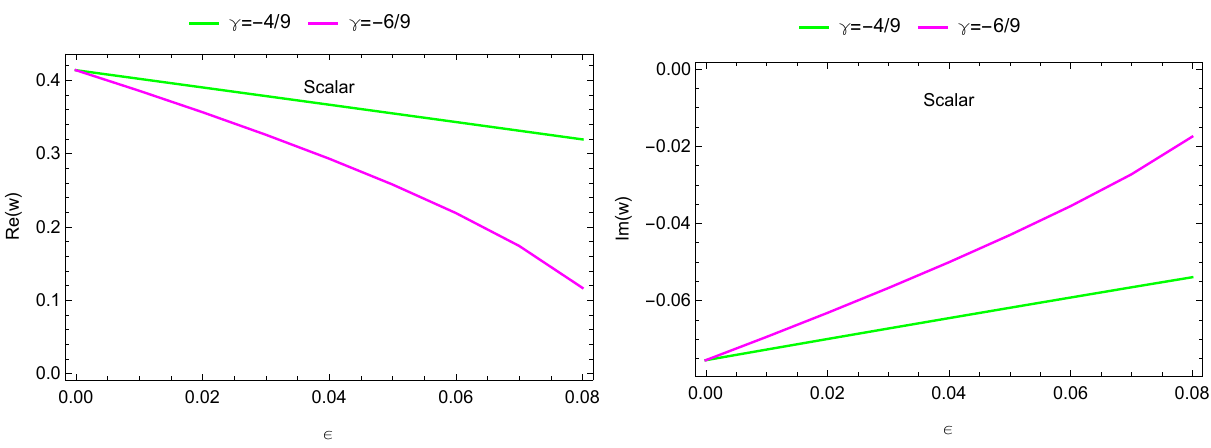} }}
    \caption{The real (left) and imaginary (right) parts of the  quazinormal frequencies (Tables \ref{taba1} and \ref{taba2}) vs $\epsilon$.}
    \label{fig4a}
\end{figure}

\begin{figure}
    \centering
{{\includegraphics[width=16cm]{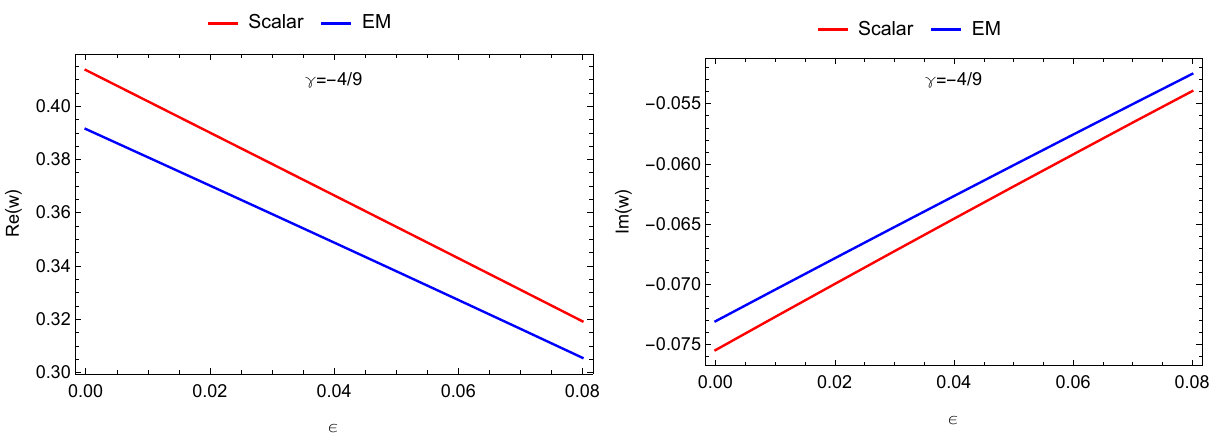} }}
    \caption{The real (left) and imaginary (right) parts of the  quazinormal frequencies (Tables \ref{taba1} and \ref{taba2}) vs $\epsilon$.}
    \label{fig4}
\end{figure}

\begin{figure}
    \centering
{{\includegraphics[width=16cm]{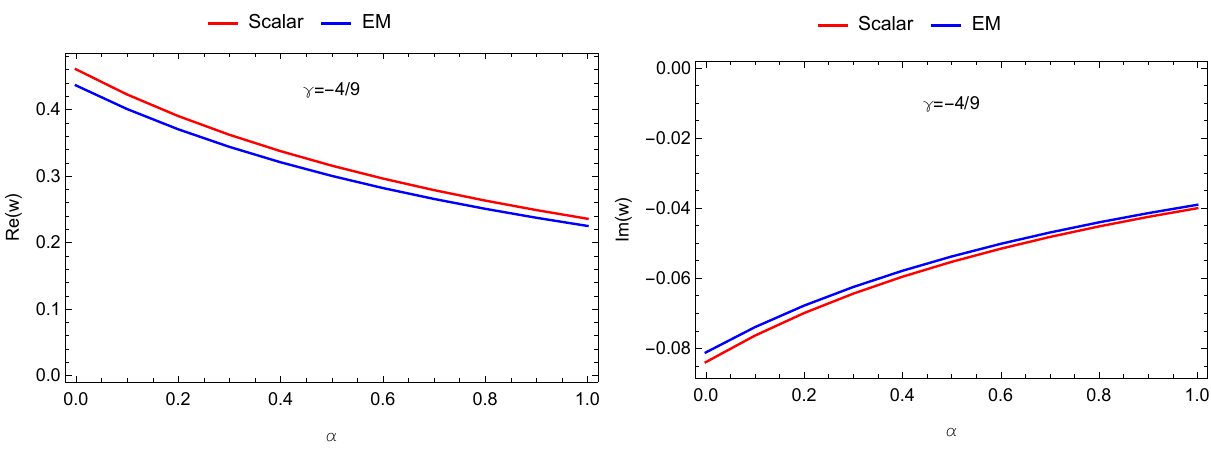} }}
    \caption{The real (left) and imaginary (right) parts of the  quazinormal frequencies (Tables \ref{taba1} and  \ref{taba2}) vs $\alpha$.}
    \label{fig5}
\end{figure}

\begin{figure}
    \centering
{{\includegraphics{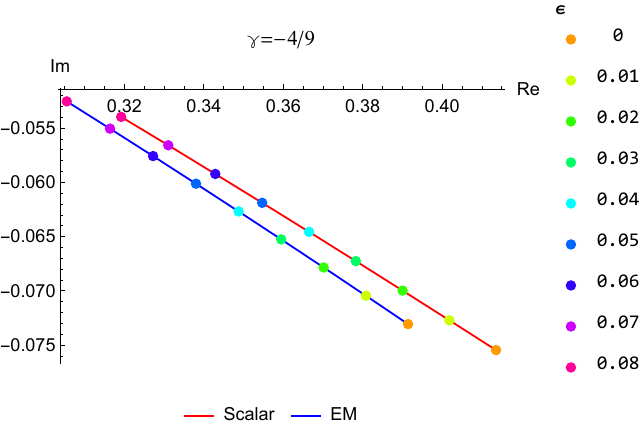} }}
    \caption{The complex frequency plane for both scalar and EM perturbations, which shows the behavior of quasinormal frequencies, Table \ref{taba1} and \ref{taba2}.}
    \label{fig6}
\end{figure}

\begin{figure}
    \centering
{{\includegraphics{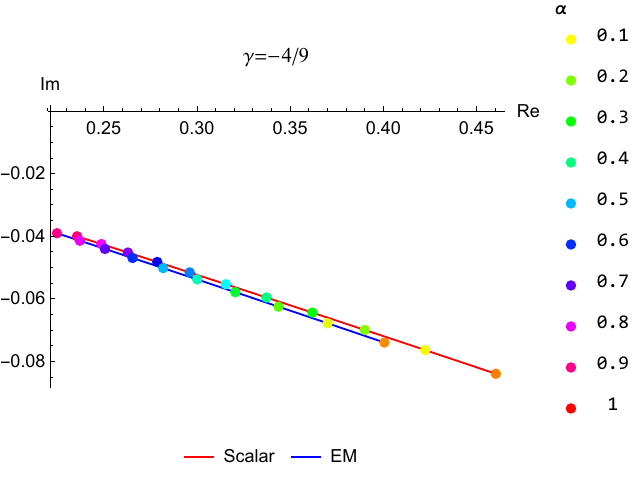} }}
    \caption{The complex frequency plane for both scalar and EM perturbations, which shows the behavior of quasinormal frequencies, Table \ref{taba1} and \ref{taba2}.}
    \label{fig7}
\end{figure}

This section aims is to investigate the propagation of a massless scalar field and Maxwell field in QMOG geometry. In general, BHs can be perturbed by adding Klein-Gordon or Dirac fields to the spacetime background or by perturbing spacetime itself. In BH perturbation theory, the external field interacts with the BH mainly through the scalar QNMs \cite{sc6}, the electromagnetic QNMs \cite{em7} and gravitational QNMs \cite{gr8}. QNM research is used to analyse the stability of BHs, which are particularly important for characterising gravitational wave signals observed by LIGO and VIRGO \cite{ligo1}. The perturbation of the QMOG metric (\ref{M1}) can be reduced to the following like-Schrödinger equation
\cite{brill1957}:
\begin{equation}
\frac{d^{2}\Psi}{dr_{\ast }^{2}}+\left( \omega^{2}-V\right) \Psi=0,  \label{s3}
\end{equation}%
where $r_{\ast }$  is the tortoise coordinate defined as $\frac{d}{dr_{\ast }}
=f\frac{d}{dr}$, and $V$ is the potential given by:
\begin{equation}
V(r
)=\frac{l\left( l+1\right) }{r^{2}}f+\frac{(1-s^{2})}{r} ff^{\prime }\label{pot1}
\end{equation}
where for $s = 0$ is the spin of the perturbing scalar field and $s = 1$ for Maxwell perturbing field.
We make plots of the potentials to investigate the characteristics of the BH potentials (\ref{pot1}) in the S-V-T modified gravity. One can gain a preliminary understanding of the GFs and QNMs by analysing the potential's behaviour. \\ Figure \ref{fig2}, represents the dependence of the scalar potential with respect to the multipole moment $l$ in the left panel and the quintessence parameter $\epsilon$ on the right panel. According to Fig. \ref{fig2}, the peaks of scalar potentials lower as $\epsilon$ increases, indicating that quintessence lowers their peaks. Similarly, Fig. \ref{fig3} shows that potentials are lowered as the MOG parameter $\alpha$ rises however the peaks are shifting to the right.  According to this analysis, both parameters $\epsilon$ and $\alpha$ have comparable effects on the potential behaviour. As a result, their effects on QNMs may be similar.  
\\The following step is to obtain the QNM frequencies. In order to compute the quasinormal frequencies efficiently, we will use the 6th WKB approximation \cite{Konoplya} that was introduced in Ref.  \cite{Iyer}. To investigate the effect of parameters $\epsilon$ and $\alpha$ on quasinormal frequencies, we focus on the fundamental mode with $l = 2$ and $n = 0$ because gravitational waves dominate. In the 6th order WKB approximation, the complex frequency formula takes the form \cite{Konoplya} \begin{equation}
i\frac{\left( \omega ^{2}-V_{max}\right) }{\sqrt{-2V_{max}^{\prime \prime }}}%
-\sum\limits_{i=2}^{6}\Lambda _{i}=n+\frac{1}{2} \label{fr2}
\end{equation}%
where $V_{max}$  is the maximum of the effective potential $V(r)$, $%
V_{max}^{\prime \prime }=\left. \frac{d^{2}V(r_{\ast })}{dr_{\ast }^{2}}%
\right\vert _{r_{\ast }=r_{max}}$ , $r_{max}$ is the position of the maximum  value
of the potential, and the WKB corrections $\Lambda _{i}$ given in Refs. \cite{Iyer,Konoplya}. The quazinormal frequencies of the QMOG BH are tabulated in Tables \ref{taba1} and \ref{taba2} using the  following model parameters: $M=1$, $l=2$, and overtone number $n = 0$. It is worth noting that, when $n<l$, the WKB method produces reliable results, and for higher $l$, its accuracy increases \cite{t33,t34,t32,rastal}.
\\Tables \ref{taba1} and \ref{taba2} show that the imaginary part is negative in all quazinormal frequency modes, indicating that the QMOG BH returns to steady state after perturbation. To see how the quintessence and the MOG parameters affect the QNMs spectrum, the real and imaginary quazinormal frequencies were plotted against these  parameters as shown in Figs. \ref{fig4a}, \ref{fig4} and \ref{fig5}. 
First, to study the impact of the quintessence's state parameter, we plot the quazinormal frequencies vs $\epsilon$ for various $\gamma$ values, as shown in Fig. \ref{fig4a}.  When $\gamma$ increases, the real component falls and the imaginary component increases. To study the effect of the quintessence parameter $\epsilon$, Figure \ref{fig4} depicts a plot of the real and imaginary parts of complex frequencies vs $\epsilon$. The real QNMs or oscillation frequencies decrease dramatically as $\epsilon$ increases. In contrast, as $\epsilon$ increases, the imaginary part of quasinormal frequencies rises, indicating a lower damping rate. Finally, to investigate the impact of the MOG parameter, we plot the quazinormal frequencies against $\alpha$, as shown in Fig. \ref{fig5}. The MOG parameter has a similar effect as the quintessence parameter $\epsilon$. However, $\epsilon$ exhibits virtually linear volatility, while $\alpha$ does not. Figures \ref{fig6} and \ref{fig7} depict the quasinormal frequencies in a complex plane at $\gamma=-4/9$. The model parameters $\epsilon$ and $\alpha$ have identical effects on the quizinormal frequencies.
\begin{center}
    \begin{tabular}{|c|c|c|c|c|c|} \hline 
          \multicolumn{3}{|c|}{$l=2$,\hspace{0.5cm}$n=0$,\hspace{0.5cm}$\alpha=0.2$}&\multicolumn{3}{|c|}{$l=2$,\hspace{0.5cm}$n=0$,\hspace{0.5cm}$\epsilon=0.02$}
\\ \hline
   $\epsilon$ & $\omega \left( \gamma=-4/9\right) $ & $\omega \left( \gamma=-6/9\right) $&$\alpha$ & $\omega \left( \gamma=-4/9\right) $ & $\omega \left( \gamma=-6/9\right) $
\\ \hline  
$0$ & $0.41356-0.075464i$ & $0.41356-0.075464i$ & $0.1$ & $0.422486-0.076412i
$ & $0.393032-0.070522i$ \\ 
$0.01$ & $0.40180-0.072711i$ & $0.38552-0.069431i$ & $0.2$ & $%
0.390048-0.069977i$ & $0.35628-0.0632149i$ \\ 
$0.02$ & $0.39004-0.069977i$ & $0.35628-0.063214i$ & $0.3$ & $%
0.361938-0.064435i$ & $0.323665-0.056775i$ \\ 
$0.03$ & $0.37827-0.067261i$ & $0.32554-0.056782i$ & $0.4$ & $%
0.337316-0.059610i$ & $0.294335-0.051026i$ \\ 
$0.04$ & $0.36649-0.064564i$ & $0.29289-0.050085i$ & $0.5$ & $%
0.315548-0.055369i$ & $0.267636-0.045834i$ \\ 
$0.05$ & $0.35469-0.061885i$ & $0.25769-0.043050i$ & $0.6$ & $%
0.296148-0.051611i$ & $0.243057-0.041096i$ \\ 
$0.06$ & $0.34288-0.059226i$ & $0.21881-0.035548i$ & $0.7$ & $%
0.278737-0.048257i$ & $0.220184-0.036728i$ \\ 
$0.07$ & $0.331046-0.05658i$ & $0.17397-0.027312i$ & $0.8$ & $%
0.263012-0.045245i$ & $0.198669-0.032664i$ \\ 
$0.08$ & $0.31918-0.053965i$ & $0.11674-0.017555i$ & $0.9$ & $%
0.248731-0.042525i$ & $0.178206-0.028847i$ \\ 
$0.09$ & $0.30729-0.051364i$ & $0.006293-0.04425i$ & $1$ & $%
0.235696-0.040055i$ & $0.158505-0.025226i$
\\ 
\hline 
\end{tabular}
\captionof{table}{The QNM of QMOG BH for  massless scalar perturbation. Here, $M=1$.} \label{taba1}
\end{center}

\begin{center}
    \begin{tabular}{|c|c|c|c|c|c|} \hline 
          \multicolumn{3}{|c|}{$l=2$,\hspace{0.5cm}$n=0$,\hspace{0.5cm}$\alpha=0.2$}&\multicolumn{3}{|c|}{$l=2$,\hspace{0.5cm}$n=0$,\hspace{0.5cm}$\epsilon=0.02$}
\\ \hline
   $\epsilon$ & $\omega \left( \gamma=-4/9\right) $ & $\omega \left( \gamma=-6/9\right) $&$\alpha$ & $\omega \left( \gamma=-4/9\right) $ & $\omega \left( \gamma=-6/9\right) $
\\ \hline  
$0$ & $0.391413-0.073068i$ & $0.391413-0.073068i$ & $0.1$ & $%
0.400599-0.073997i$ & $0.374767-0.068098i$ \\ 
$0.01$ & $0.380801-0.070450i$ & $0.366558-0.067150i$ & $0.2$ & $%
0.370157-0.067846i$ & $0.340398-0.061094i$ \\ 
$0.02$ & $0.370157-0.067846i$ & $0.340398-0.061094i$ & $0.3$ & $%
0.343755-0.062538i$ & $0.309871-0.054911i$ \\ 
$0.03$ & $0.359479-0.065256i$ & $0.312636-0.054868i$ & $0.4$ & $%
0.320611-0.057910i$ & $0.282391-0.049383i$ \\ 
$0.04$ & $0.348765-0.062682i$ & $0.282838-0.048426i$ & $0.5$ & $%
0.300134-0.053835i$ & $0.257347-0.044386i$ \\ 
$0.05$ & $0.338012-0.060122i$ & $0.250327-0.041695i$ & $0.6$ & $%
0.281873-0.050220i$ & $0.234261-0.039821i$ \\ 
$0.06$ & $0.327217-0.057577i$ & $0.213935-0.034543i$ & $0.7$ & $%
0.265473-0.046990i$ & $0.21274-0.03561i$ \\ 
$0.07$ & $0.316377-0.055048i$ & $0.1713-0.026693i$ & $0.8$ & $%
0.250652-0.044085i$ & $0.192457-0.031696i$ \\ 
$0.08$ & $0.305488-0.052535i$ & $0.115862-0.017323i$ & $0.9$ & $%
0.237184-0.041458i$ & $0.173118-0.028017i$ \\ 
$0.09$ & $0.294546-0.050037i$ & $0.006309-0.04430i$ & 1 & $0.224885-0.039071i
$ & $0.154443-0.024529i$%
\\ 
\hline 
\end{tabular}
\captionof{table}{The QNM of QMOG BH for  EM perturbation.  Here, $M=1$.} \label{taba2}
\end{center}
Furthermore, the authors of \cite{qu1,qu2,qu3,qu4,qu7,qu5,qu6,qu8} studied the QNMs of multiple BHs surrounded by quintessence using the WKB approximation approach and the continuous fraction method to determine the impact of a quintessence scalar field. Their studies showed that the existence of a quintessence field alters quasinormal frequencies, so that as the quintessence parameters increase, the oscillations of the scalar and Maxwell fields damp more slowly. In our study, we used the 6th order WKB method to analyse massless scalar and electromagnetic perturbations QNMs in QMOG BH, and we discovered that both frequency and decay rate decrease as the quintessence parameters increase.


\section{Greybody factors}

\begin{figure}
    \centering
{{\includegraphics[width=7.5cm]{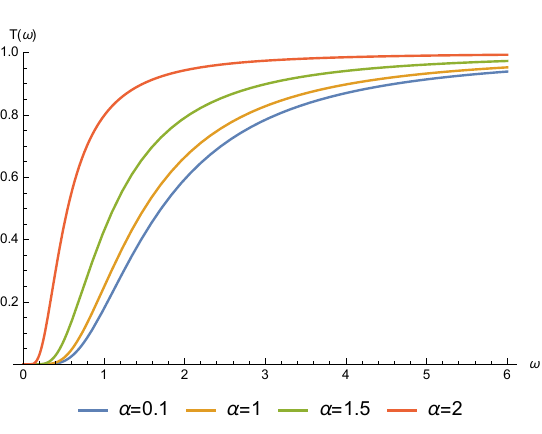} }}\qquad
    {{\includegraphics[width=7.5cm]{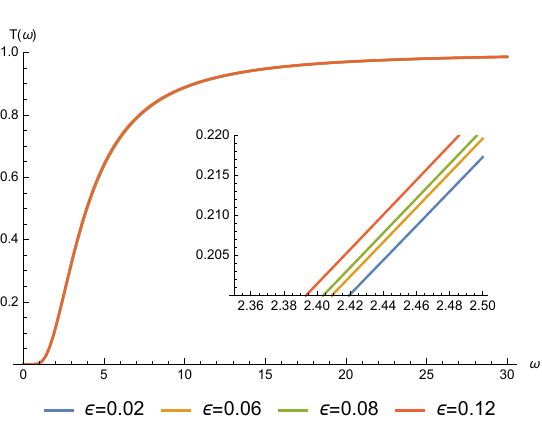}}} \caption{The greybody bounding of scalar massless of QMOG BH  Eq. (\ref{gf1}) and different values of $\alpha$ (left) and $\epsilon$ (right).}
\label{Fig4}
\end{figure}

The purpose of this section is to analyse the GF for the QMOG BH. The outer curvature and gravitational barrier around a BH affect the radiation spectrum it emits, also known as Hawking radiation.  The impacted spectrum is referred to as the GF because it combines the effect of the surrounding spacetime, which serves as a potential radiation barrier, with the black body spectrum. This results in a deviation from the black-body radiation spectrum as seen by the asymptotic observer. To summarise, the GF is the difference between the spectra of asymptotic and black-body radiations \cite{qn42,qn99}. Many theoretical investigations of GFs, or transmission probabilities of Hawking radiation for different gravity models and BH solutions, may be found in \cite{qn32,qn36,qn37,qn38,qn39,qn40,qn41,qn42,qn43,qn44,qn45,qn46} using various approaches. We will utilise the general semi-analytic bounds approach to determine the bounds of the QMOG BH's GF. Using this method, the  bounds of the GF can be obtained:   \cite{qn41,qn44,qn54} 
\begin{equation}
T\left( w\right) \geq \sec h^{2}\left( \frac{1}{2w}%
\int_{r_{h}}^{+\infty }V_{eff}dr_{\ast }\right) ,  \label{is8}
\end{equation}
Considering the effective potential of massless scalar given in Eq. (\ref{pot1}) we can obtain the bounds of the GF of QMOG BH. 
 Therefore, Eq. (\ref{is8}) becomes
\begin{equation}
T\left( w\right) \geq \sec h^{2}\left( \frac{1}{2\omega }
\int_{r_{h}}^{\infty }\left( \frac{l\left( l+1\right) }{r^{2}}+\frac{f^{\prime }}{r}\right) dr \right) .\label{in10}
\end{equation}
We manage to obtain the analytical solution of the bounds of the GF given Eq. (\ref{in10}) as 
\begin{equation}
    T\left( w\right) \geq \sec h^{2}\left[ \frac{1}{\omega }\left( -\frac{l(l+1)%
}{r_{h}}+\frac{\left( 1+\alpha \right) M}{r_{h}^{2}}-\frac{\epsilon r_{h}^{1-3\gamma
}\left( 1+3\gamma \right) \left( 1+\alpha \right) }{\left( 2+3\gamma \right)
r_{h}^{3}}-\frac{\alpha M^{2}\left( 1+\alpha \right) }{3r_{h}^{3}}\right)
\right]. \label{gf1}
\end{equation}
We plot the transmission
coefficients (\ref{gf1}) versus $\omega$ for different values of the MOG and quintessence parameters in Fig. \ref{Fig4}.  
It is clear that for higher values of $\alpha$, the value of transmission
coefficients  significantly increases, implies more thermal radiation to reach the observer at spatial infinity. We also observe similar trend for the parameter $\epsilon$ as seen in the right panel of Fig. \ref{Fig4}. \\ According to WKB scattering, the fields near the horizon and infinity can have the following asymptotic forms: \cite{gf09}: \begin{equation}
\psi \left( x\right) =\left\{ 
\begin{array}{cc}
T\left( \omega \right) e^{-i\omega x}, & x\rightarrow -\infty  \\ 
e^{-i\omega x}+R\left( \omega \right) e^{i\omega x} & x\rightarrow \infty 
\end{array}%
\right. 
\end{equation}%
where $T\left( \omega \right) $ and $R\left( \omega \right) $ represent
transmission and reflection  coefficients respectively. Recall that, the
condition of the conservation of probability requires that $\left\vert
T\right\vert ^{2}+$ $\left\vert R\right\vert ^{2}=1.$ The two coefficients
can be obtained as \cite{gf09} 
\begin{equation}
\left\vert T\right\vert ^{2}=\frac{1}{1+e^{2\pi i\delta }},
\end{equation}%
\begin{equation}
\left\vert R\right\vert ^{2}=\frac{1}{1+e^{-2\pi i\delta }},
\end{equation}%
where $\delta $ can be obtained from Eq. (\ref{fr2}). 

\section{Plasma effects to the Shadow of QMOG BH}

The aim of this section is to show how the plasma background as well as the model parameters affect the
QMOG BH shadow \cite{g1}. The main reason for focusing on this question is due to the fact that most BHs are surrounded by a medium which changes their geodesic \cite{plazma,pz2,pz3}.
To begin, consider a plasma with a refractive index of $n=n(x^{i},\omega ),$
where $\omega $, denotes photon frequency. The plasma background surrounding the QMOG BH alters the Hamiltonian by introducing new terms into the geodesic equations. As a result, these changes influence particle trajectories and have a clear frequency dependence.  Assuming the effective energy of the photons inside the plasma
medium as $\hbar \omega =p_{\mu }u^{\mu }$, then the refraction index of the plasma is calculated as a function of the photon four-momentum as \cite{62}:\begin{equation}
n^{2}=1+\frac{p_{\mu }p^{\mu }}{\left( p_{\mu }u^{\mu }\right) ^{2}} ,
\end{equation} where for vacuum we have $n=1$. Assuming that the refractive index has the general form, one can introduce a specific form for the plasma frequency for analytic processing\begin{equation}
n^{2}=1-\left( \frac{\omega _{p}}{\omega }\right) ^{2},
\end{equation}%
 where $\omega _{p}$ denotes the plasma frequency. After that, the refractive
index of plasma is calculated in radial power law form, as described by \cite{63,64}
\begin{equation}
n\left( r\right) =\sqrt{1-\frac{\rho }{r}},
\end{equation}%
where the plasma background parameter $\rho \geq 0.$ 

\begin{figure}
    \centering
    \includegraphics{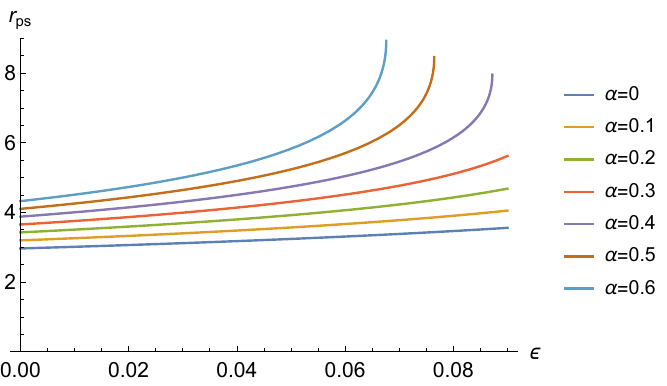}
    \caption{The variation of the
 $r_{ps}$ vs $\epsilon$ for different values of $\alpha$. Here, $\rho=0.6$.}
    \label{figPS}
\end{figure}

\begin{figure}
    \centering
    \includegraphics{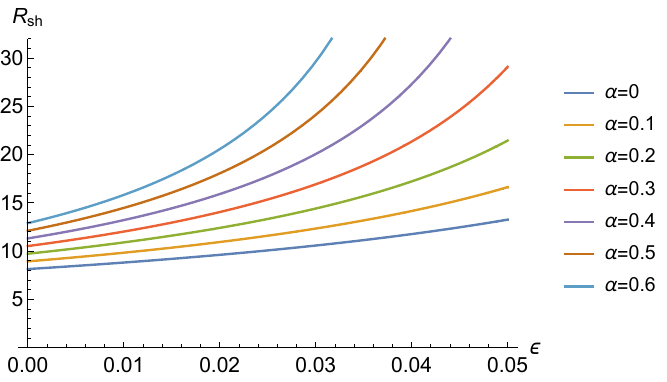}
    \caption{The variation of the
$R_{sh}$ vs $\epsilon$ for different values of  $\alpha$. Here, $\rho=0.6$.}
    \label{figSH}
\end{figure}

The Hamilton-Jacobi equation in the plasma background \cite{62} is:%
\begin{equation}
\frac{\partial \mathcal{S}}{\partial \sigma }=-\frac{1}{2}\left( g^{\mu \nu
}p_{\mu }p_{\nu }+\left( 1-n^{2}\right) \left( p_{0}\sqrt{-g^{00}}\right)
^{2}\right) ,
\end{equation}%
where $\mathcal{S}$ is the Jacobi action.  We can obtain the equations of
motion namely, 
\begin{equation}
\overset{\cdot }{t}=\frac{n^{2}E}{f},\qquad \frac{d\phi }{d\sigma }=-\frac{%
\ell }{r^{2}\sin ^{2}\theta },
\end{equation}%
\begin{equation}
r^{2}\overset{\cdot }{r}=\pm \sqrt{n^{2}r^{4}E^{2}-\left( \mathcal{K}+\ell
^{2}\right) r^{2}f},  \label{R1}
\end{equation}%
\[
r^{2}\overset{\cdot }{\theta }=\pm \sqrt{\mathcal{K}-\ell ^{2}\cot \theta },
\]%
where $\mathcal{K}$ is the Carter separation constant,  $E$ and $\ell =r^{2}\sin
^{2}\theta \overset{\cdot }{\phi }$ are the Energy and the angular momentum.  Defining  impact parameters as $\eta =%
\frac{\mathcal{K}}{E^{2}}$ and $\zeta =$ $\frac{\ell }{E}$.  It is well
known that shadow casts can be obtained by using unstable null circular
orbits. Therefore, we have to re-express the radial null geodesic equation
as 
\begin{equation}
\left( \frac{dr}{d\sigma }\right) ^{2}+V_{eff}=0,
\end{equation}%
where%
\begin{equation}
V_{eff}\left( r\right) =\frac{f}{r^{2}}\left( \mathcal{K}+\ell ^{2}\right)
-n^{2}E^{2}.
\end{equation}%
Furthermore, the effective potential of the Schwarzschild scenario coincides
perfectly with the zero limit of $\alpha $ and $\epsilon$. Circular orbits
correspond to the maximum effective potential, and the unstable photons must
meet the following requirements:%
\begin{equation}
V_{eff}\left( r\right) \left\vert _{r=r_{ps}}\right. =0,V_{eff}^{\prime
}\left( r\right) \left\vert _{r=r_{ps}}\right. =0, \label{R3}
\end{equation}
where $r_{ps}$ is the radius of the unstable photon sphere. Using  Eq. (\ref
{R3}), we can obtain the equation for   $r_{ps}$ namely%
\begin{equation}
n\left( r_{ps}\right) r_{ps}f^{\prime }\left( r_{ps}\right) -2n\left(
r_{ps}\right) f\left( r_{ps}\right) -2r_{ps}n^{\prime }\left( r_{ps}\right)
f\left( r_{ps}\right) =0.  \label{rps1}
\end{equation}%
and then, with the help of impact parameters we obtain%
\begin{equation}
\eta +\zeta ^{2}=r_{ps}^{2}\frac{n^{2}\left( r_{ps}\right) }{f\left(
r_{ps}\right) }=R_{sh}^{2},
\end{equation}%
where $R_{sh}$ is the shadow radius. Analytically, solving Eq. (\ref{rps1})
is extremely difficult. Consequently, we present numerical analyses and plots showing the effect of BH parameters on the photon radius of mass particles and the shadow radius of BH. The numerical results for the photon sphere and the shadow radius are
presented in Table \ref{taba3}. According to the Table, as the plasma background parameter increases, the shadow radius decreases. Furthermore, it appears that as $\alpha$ and $\epsilon$ increase, so do the radii ($r_{ps}$ and $R_{sh}$). Despite this, when $\alpha$ remains constant, increasing $\epsilon$ causes more shadow radius growth. Figure \ref{figPS} shows the  behaviour of the size of photon sphere and how the MOG parameter $\alpha$ and quintessence parameter $\epsilon$ influence the radius $r_{ps}$ of massive particles surrounding the QMOG BH. One can see that $r_{ps}$ increases as $\alpha$ and $\epsilon$ increase. Similarly, Fig. \ref{figSH} shows the shadow radius. The same pattern was observed with an increase in $\alpha$ and $\epsilon$.
\begin{center}
    \begin{tabular}{|c|c|c c|c c|c c|} \hline 
 & &  \multicolumn{2}{|c|}{$\epsilon=0.01$} &   \multicolumn{2}{|c|}{$\epsilon=0.02$}  &  \multicolumn{2}{|c|}{$\epsilon=0.03$}   \\ \hline
 $\rho$ &$\alpha $ & $r_{ps}/M$ & $R_{sh}/M$ & $r_{ps}/M$ & $R_{sh}/M$ & $r_{ps}/M$ & 
$R_{sh}/M$ \\ \hline
&$0.1$ & $3.259$ & $10.543$ & $3.326$ & $11.710$ & $3.399$ & $12.063$ \\ 
&$0.2$ & $3.505$ & $11.623$ & $3.593$ & $13.204$ & $3.690$ & $15.309$ \\  $0.2$&
$0.3$ & $3.753$ & $12.760$ & $3.865$ & $14.878$ & $3.992$ & $17.887$ \\ &
$0.4$ & $4.002$ & $13.963$ & $4.145$ & $16.777$ & $4.309$ & $21.104$ \\ &
$0.5$ & $4.254$ & $15.242$ & $4.432$ & $18.966$ & $4.643$ & $25.270$ \\ &
$0.6$ & $4.509$ & $16.609$ & $4.729$ & $21.530$ & $4.999$ & $30.932$ \\ \hline & 
$0.1$ & $3.222$ & $10.314$ & $3.290$ & $11.455$ & $3.365$ & $12.894$ \\ &
$0.2$ & $3.469$ & $11.389$ & $3.558$ & $12.938$ & $3.657$ & $14.999$ \\  $0.4$&
$0.3$ & $3.717$ & $12.521$ & $3.831$ & $14.598$ & $3.961$ & $17.549$ \\ &
$0.4$ & $3.966$ & $13.718$ & $4.111$ & $16.482$ & $4.279$ & $20.729$ \\ &
$0.5$ & $4.218$ & $14.990$ & $4.400$ & $18.652$ & $4.615$ & $24.847$ \\ &
$0.6$ & $4.473$ & $16.350$ & $4.698$ & $21.193$ & $4.973$ & $30.441$ \\ \hline  &
$0.1$ & $3.181$ & $10.078$ & $3.252$ & $11.193$ & $3.329$ & $12.599$ \\ &
$0.2$ & $3.428$ & $11.149$ & $3.520$ & $12.664$ & $3.622$ & $14.682$ \\  $0.6$ &
$0.3$ & $3.677$ & $12.276$ & $3.795$ & $14.312$ & $3.927$ & $17.203$ \\ &
$0.4$ & $3.927$ & $13.468$ & $4.075$ & $16.180$ & $4.247$ & $20.347$ \\ &
$0.5$ & $4.180$ & $14.734$ & $4.3653$ & $18.331$ & $4.585$ & $24.416$ \\ &
$0.6$ & $4.435$ & $16.087$ & $4.664$ & $20.850$ & $4.946$ & $29.940$%
\\ 
\hline 
\end{tabular}
\captionof{table}{The numerical values of $r_{ps}$ and $R_{sh}$ with three different values of
plasma background parameter $\rho=0.2,0.4$ and $0.6$ with $\gamma=-2/3$.} \label{taba3}
\end{center}

Now we'll look at the visualisation of the BH shadow, which is a geometrical quantity on a celestial plane along the coordinates $X$ and $Y$. As a result, the celestial coordinates are given by \cite{Vazquez}:

\begin{equation}
X=\lim_{r_{0}\rightarrow \infty }\left( -r_{0}\sin \theta _{0}\left. \frac{%
d\phi }{dr}\right\vert _{r_{0},\theta _{0}}\right) ,  \label{x11}
\end{equation}%
\begin{equation}
Y=\lim_{r_{0}\rightarrow \infty }\left( r_{0}\left. \frac{d\theta }{dr}%
\right\vert _{r_{0},\theta _{0}}\right) ,  \label{y11}
\end{equation}%

where $r_0$ denotes the distance between the BH and the observer. To be more specific, we consider null geodesic motion in the equatorial plane $\theta=0$, which leads to $X=-\xi$ and $Y=\pm \sqrt{\eta}$. Thus, this result has a two-dimensional geometry governed by the shadow radius, which is expressed as follows:
\begin{equation}
X^{2}+Y^{2}=R_{s}^{2}=\eta +\zeta ^{2},  \label{xy1}
\end{equation}
which is nothing more than the shadow radius $R_{sh}$ in celestial coordinates. It should be noted that the shadow shape for static BHs is a circle with a radius of $R_{sh}$. The following Figs. \ref{Fig12}- \ref{Fig14} illustrate the relationship between the shadow radius, $R_{sh}$, and the parameters $\epsilon$, $\alpha$, and plasma. The fact that increasing $\epsilon$ has a greater impact on the shadow radius than increasing $\alpha$ is particularly intriguing, indicating that the quintessence parameter $\epsilon$ has a greater impact than the MOG parameter $\alpha$.
\begin{figure}
    \centering
{{\includegraphics[width=7.5cm]{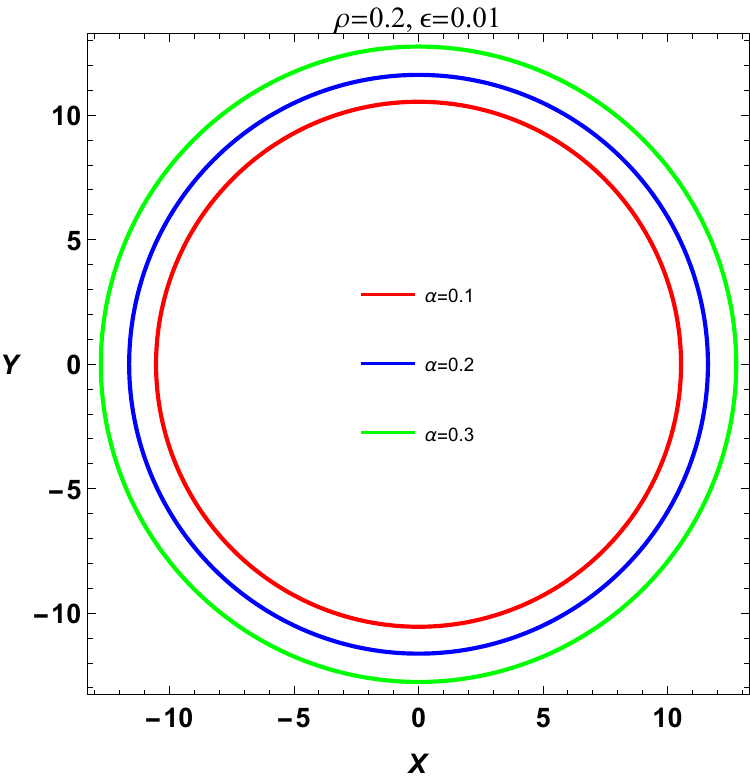} }} \qquad
    {{\includegraphics[width=7.5cm]{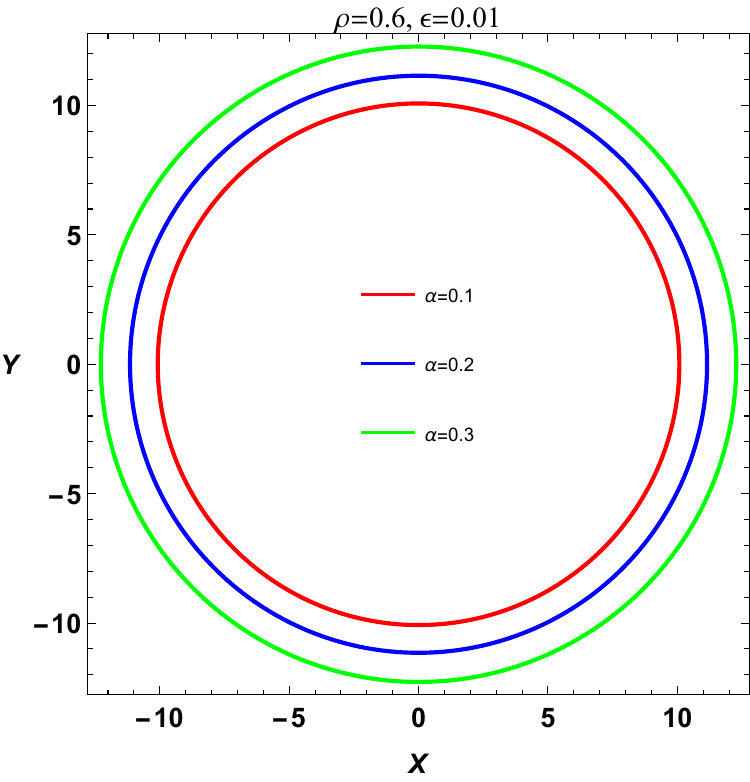}}} \caption{The shadow of the QMOG BH in a plasma background with several values of $\alpha$ parameter. Here, $\gamma=-2/3$.}
\label{Fig12}
\end{figure}
\begin{figure}
    \centering
{{\includegraphics[width=7.5cm]{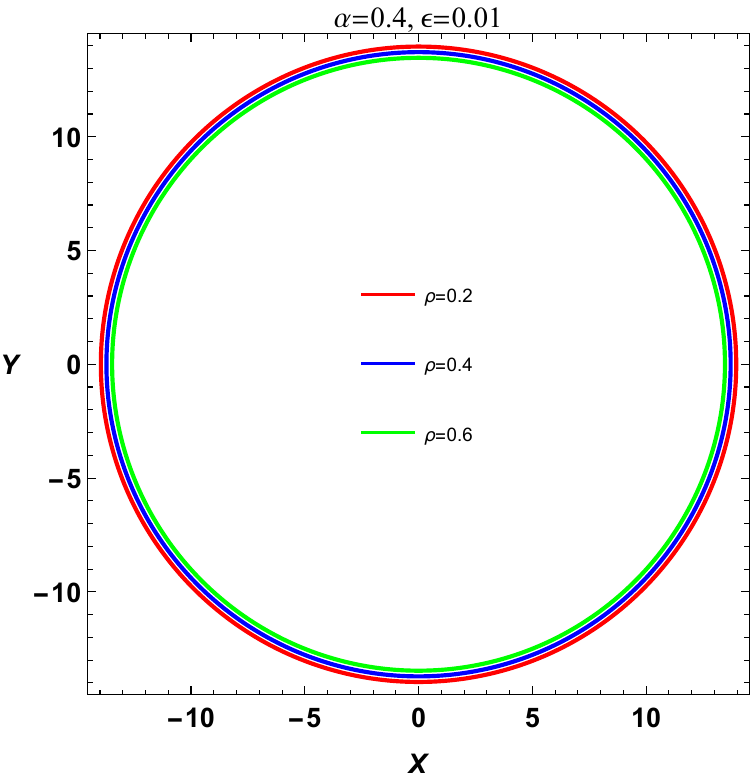} }} \qquad
    {{\includegraphics[width=7.5cm]{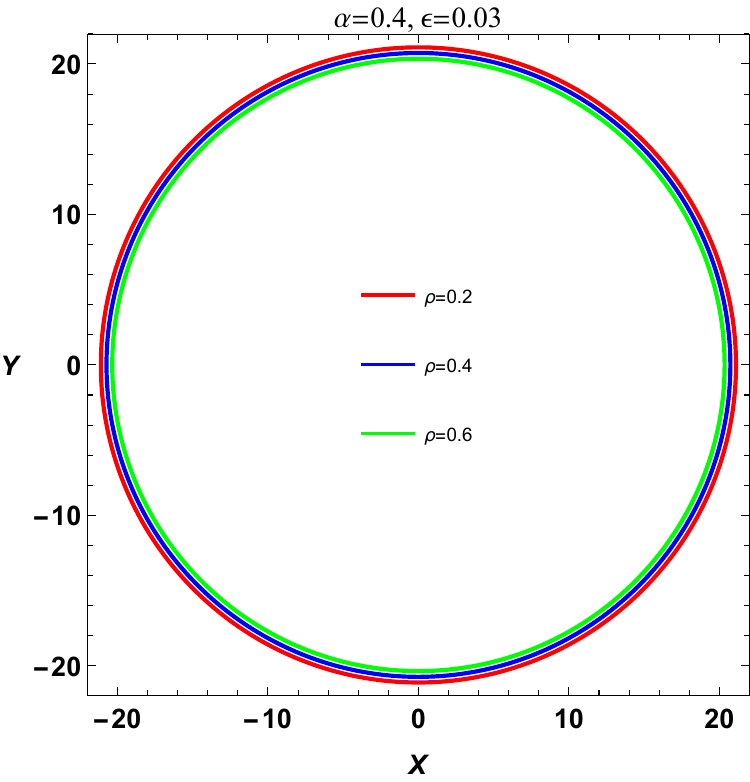}}} \caption{The shadow of the QMOG BH for different
values of plasma background parameter. Here, $\gamma=-2/3$.}
\label{Fig13}
\end{figure}\begin{figure}
    \centering
{{\includegraphics[width=7.5cm]{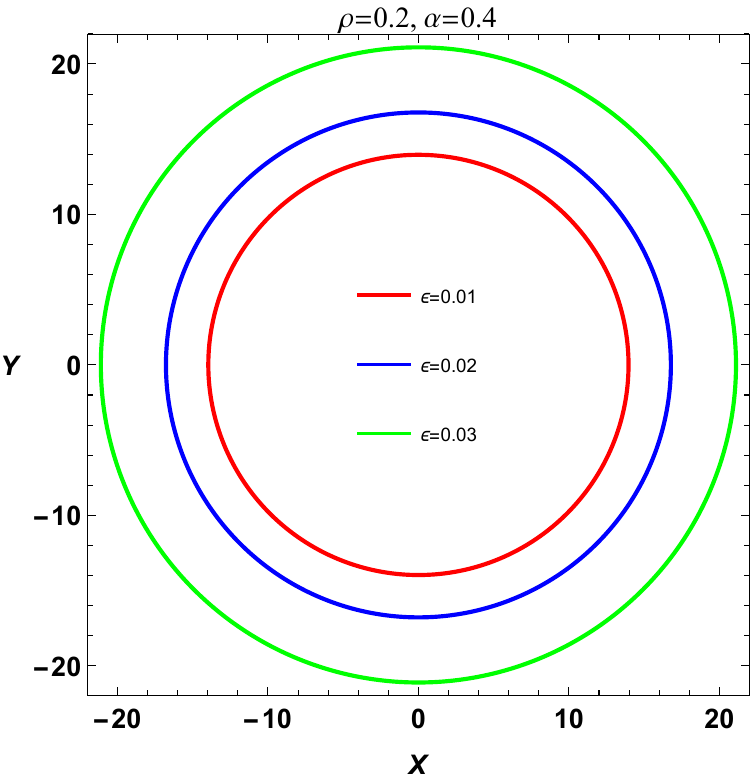} }} \qquad
    {{\includegraphics[width=7.5cm]{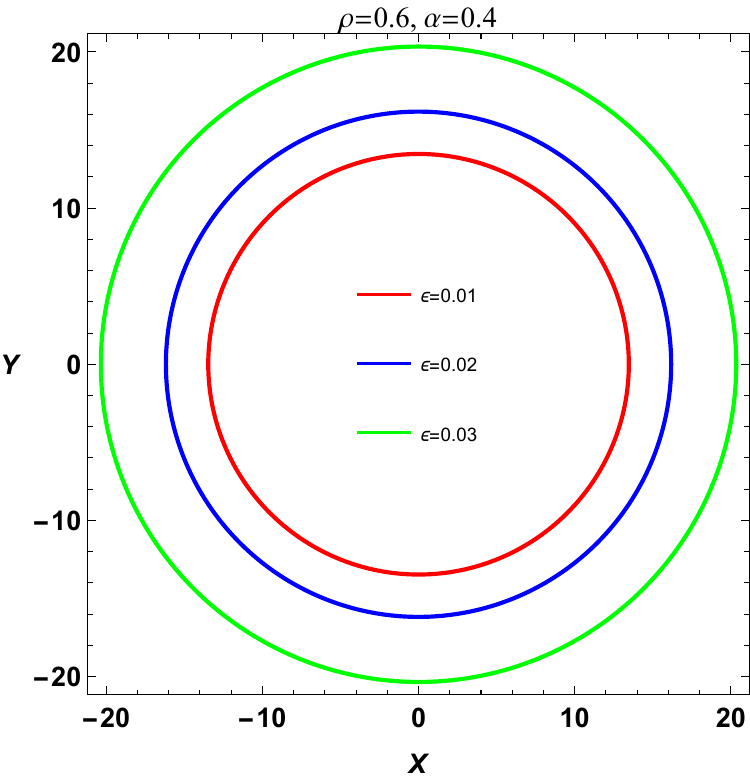}}} \caption{The shadow of the QMOG BH in a plasma background with several 
values of $\epsilon$. Here, $\gamma=-2/3$.}
\label{Fig14}
\end{figure}
\\ On the other hand, we hypothesise that the BH shadow corresponds to the high energy absorption cross section for the observer at infinity. For a spherically symmetric black hole, the absorption cross section often oscillates around a limiting constant value $(\sigma_{lim})$. The shadow, which measures the optical appearance of a black hole, is really equal to the limiting constant value of the high-energy absorption cross section. Thus, the limiting constant value, $\sigma_{lim}$, can be roughly represented as\begin{equation}
    \sigma_{lim} \approx \pi R_{sh}
\end{equation} When the black hole emits energy at a high energy, the energy emission rate is as follows:\begin{equation}
    \frac{d^2E}{d\omega dt}=\frac{2\pi^3 R_{sh}^2}{e^{\omega /T}-1}\omega^3
\end{equation} where $T$ represents the Hawking temperature.  Figure \ref{figenergy} displays the energy emission rate vs frequency $\omega$ for different MOG parameters $\alpha$. It shows that the BH's energy emission rate reaches a peak. When the MOG parameter increases, the peak moves to a higher frequency. \begin{figure}
    \centering
    \includegraphics{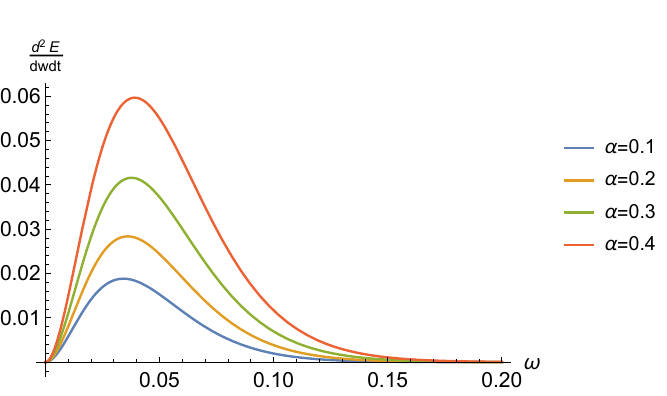}
    \caption{The behavior of energy emission rate  of the QMOG BH for various values of the MOG parameter $\alpha$.}
    \label{figenergy}
\end{figure}

\section{Conclusion}
In this work, we investigated QNMs, GFs, and shadow in plasma in relation to the possible presence of exotic forms of matter around BH type cosmic objects. We used the static spherical symmetric vacuum solution of the field equations in  the S-V-T MOG \cite{Moafat2006}. Furthermore, the QMOG BH studied in this paper possesses numerous intriguing physical and astrophysical characteristics that could lead to a deeper understanding of the physics of BHs surrounded by quintessence type exotic fluid. \\ First we computed the scalar massless field and the EM field QNMs for the QMOG BH using the 6th order WKB approximation method and investigate the dependence of the oscillation amplitude and damping on the quintessence and MOG parameters. It is observed that the real QNMs or oscillation frequencies decrease significantly as $\epsilon$ and $\alpha$ increase indicating a weak oscillation in the system.  However, as $\epsilon$ and $\alpha$ increase, the imaginary part of quasinormal frequencies rises, indicating a lower damping rate.    Next, we investigated GFs for scalar fields with varying MOG and quintessence parameters. Our analysis revealed that as $\alpha$ increases, the GF increases significantly, implying that more thermal radiation will reach the observer at spatial infinity. We see a similar pattern for the parameter $\epsilon$. \\ We looked into how the plasma background influences the QMOG BH shadow. The shadow radius decreases as the plasma background parameter increases. Furthermore, the quintessence and MOG parameters affect the BH shadow in the following ways: The radius of the shadow grows as $\epsilon$ and $\alpha$ increase. The fact that increasing $\epsilon$ has a greater impact on the shadow radius than increasing $\alpha$ is especially intriguing, implying that the quintessence parameter $\epsilon$ has a greater impact than the MOG parameter $\alpha$. \\ We considered QNMs, GFs, and shadow in plasma in relation to the possible presence of exotic forms of matter around BH type cosmic objects in this study. The findings may offer up new paths for observational testing of this type of object, as well as distinguishing between other forms of compact objects. Future research could look into these corrections for other types of BHs, such as rotating BHs.

\end{document}